# Mesoscopic Light Localization and Inverse Participation Ratio Analysis of Tissue Structural Disorder for Optical Cancer Detection

Santanu Maity, Mousa Alrubayan, Prabhakar Pradhan*

[1]*Department of Physics and Astronomy, Mississippi State University, Mississippi State, MS 39762.*

*Corresponding Authors: pp838@msstate.edu

**Abstract:** We introduce a mesoscopic physics-based framework that transforms conventional transmission optical micrographs into quantitative maps of tissue structural heterogeneity using the Inverse Participation Ratio (IPR). For the first time, to our knowledge, IPR-based light-localization analysis is applied to cancer tissue imaging to quantify nano- to submicron-scale structural alterations through spatial fluctuations in tissue mass density or refractive index. Unlike conventional morphology-based assessment, this physics-driven approach provides objective structural biomarkers from label-free or routinely stained tissue images. The method establishes a scalable, reproducible platform for quantitative computational pathology and enhanced cancer diagnosis by integrating mesoscopic optical physics with standard optical microscopy.

***Introduction:*** Mesoscopic optical physics provides a powerful framework for probing wave propagation, scattering, interference, and localization in structurally disordered media [1–4]. It is particularly sensitive to nanometer-to-submicron structural variations that may remain unresolved by conventional optical microscopy. In weakly disordered systems, optical-eigenfunction localization is governed by the magnitude of refractive-index fluctuations *dn* (where $n(r)=n_0+dn$, dn Gaussian with $<dn>=0$) and their spatial correlation decay length $l_c$. Mesoscopic light-localization parameters can therefore reveal information beyond conventional morphological assessment and provide quantitative measures of structural disorder.

Biological tissues can be treated as weakly disordered optical media, in which mass-density variations produce corresponding refractive-index fluctuations that influence light scattering, propagation, and localization. In transmission micrographs, these structural variations appear as intensity fluctuations, enabling mesoscopic optical analysis to quantify nanometer-to-submicron heterogeneity and to extract optics-physics-based structural biomarkers from conventional tissue images. This capability is important for tissue characterization and abnormality detection, such as cancer. Cancer remains a leading cause of mortality worldwide, accounting for millions of deaths annually [5–7]. Accurate early-stage detection enables timely intervention, improved disease management, and better clinical outcomes [8–11]. Although histopathological examination of biopsy specimens remains the clinical gold standard, it largely relies on pathologists' visual assessments and chemical staining, which can be time-consuming, expensive, subjective, and error-prone. These limitations create a need for quantitative optical biomarkers that objectively characterize tissue structure and support conventional pathology.

Cancer progression involves molecular reorganization, accumulation of intracellular DNA, RNA, and lipids, extracellular matrix components, and uncontrolled cellular growth. These changes increase tissue mass-density heterogeneity and cellular-level redistribution, modifying tissue spatial architecture. The resulting fluctuations in mass density and refractive index affect light propagation and localization. Structural disorder is determined by the standard deviation of dn ($\sigma(dn)$) and their spatial correlation, all of which increase with malignancy. Previous studies have shown that structural disorder increases with cancer progression, supporting structural disorder as a potential diagnostic biomarker [12–14].

Existing methods for quantifying tissue bulk disorder include phase-contrast microscopy, optical coherence tomography, and disorder fluctuations using partial-wave spectroscopy. These often require complex light-scattering experiments, specialized instrumentation, expert personnel, and nonroutine sample preparation. In contrast, conventional optical microscopy is widely used for H&E-based tissue assessment, thereby motivating the development of a computational framework to extract quantitative information on structural disorder directly from standard micrographs of stained and unstained tissues.

Here, we introduce a mesoscopic optical-physics-based technique that performs virtual numerical light experiments on images of unstained thin tissue samples. It uses the inverse participation ratio (IPR), originally developed to quantify wave propagation and localization in disordered media, to measure nanometer-to-submicron spatial disorder [15,16]. These biomarkers can objectively support conventional pathology. The standard deviation of IPR consistently increases with cancer stage across several cancer types, establishing IPR-based light-localization parameters as robust optical diagnostic biomarkers. This work extends mesoscopic light-localization theory to routine microscopy and provides a simple, objective, reproducible, scalable, and cost-effective platform for quantitative cancer detection and staging.

***Methods and Techniques:*** Cancer progression alters tissue/cell structures by rearranging and later increasing mass density of basic building blocks, such as DNA, RNA, lipids, and other macromolecules [1–3]. In this study, brightfield microscopy image contrast arises from spatial variations in tissue mass density ($\rho$), which affect the refractive index (RI), $n(r)$, and, consequently, the transmission intensity [12]. The transmitted intensity, $I_t$ *(as a tissue micrograph in Fig.1):*

$$I_t(x,y) \propto \rho(x,y) \propto n(x,y) .$$

These intensity variations enable analysis of structural disorder in label-free images and imaging of strained tissue.

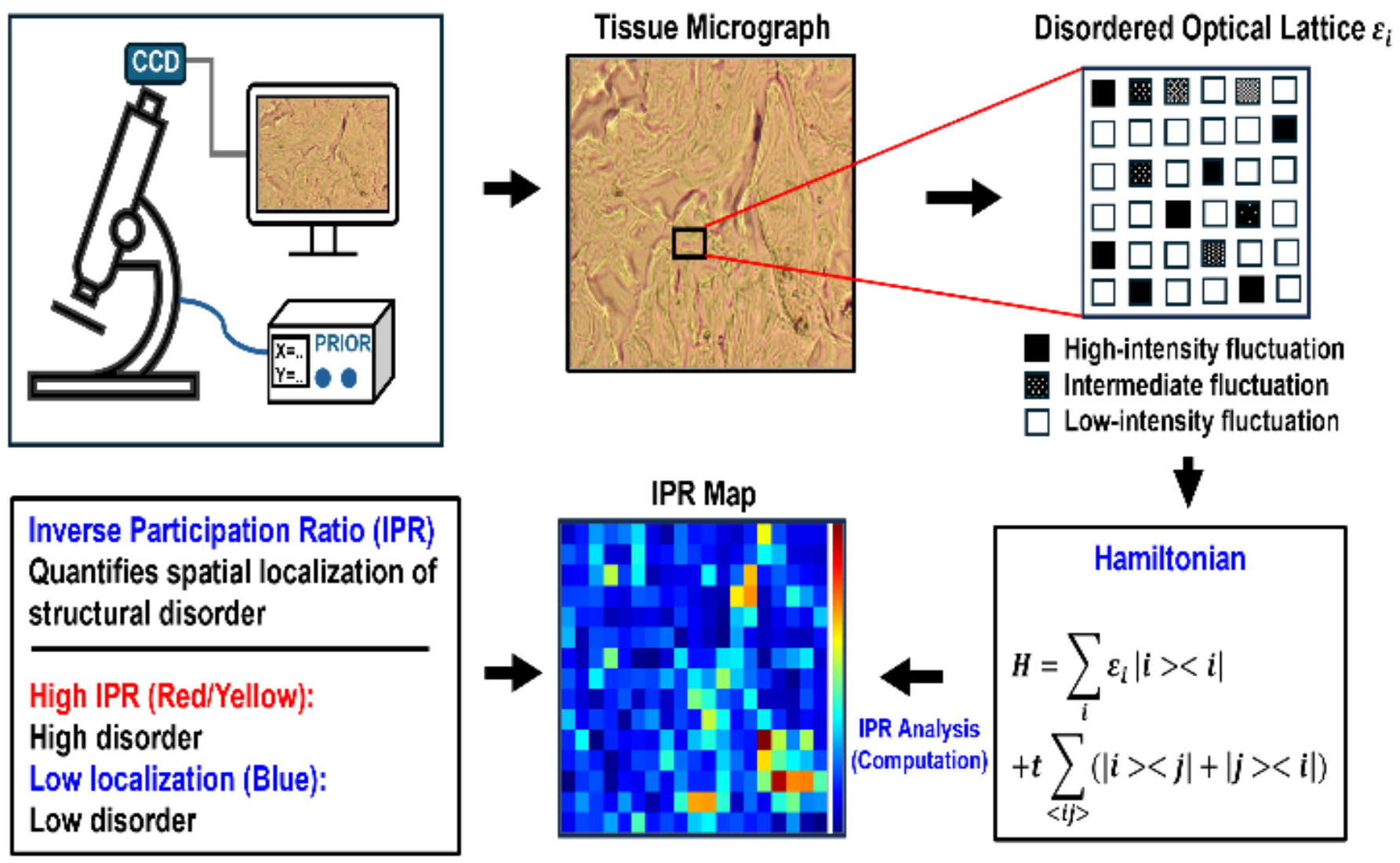

Fig.1 Flowchart for quantifying tissue structural disorder from optical micrographs using the light-localization *IPR* method. For heterogeneous media, ($L_d$) represents the total structural disorder determined by the combined effects of refractive-index STD of fluctuations *σ(dn)*, and their spatial correlation length ($l_c$).

***Structural Disorder by using IPR:*** To quantify structural disorder, we adapted the Inverse Participation Ratio (IPR), a light-localization metric from mesoscopic condensed-matter physics. The technique is shown in the schematic flowchart in Fig.1. First, we developed the technical framework for constructing a disordered optical (refractive index, RI) lattice, represented as a 2D refractive index matrix $n(x,y) = n_0 + dn(x,y)$ from optical transmission micrographs. From this, we can find that the pixel intensity measured by transmission microscopy can be directly related to the mass density present within the thin tissue voxel:

$$\frac{dI(x,y)}{\langle I\rangle} \propto \frac{d\rho(x,y)}{\langle \rho\rangle}.$$

It has also been established that, in tissue media, refractive index fluctuations are proportional to mass density fluctuations $\frac{dn(x,y)}{\langle n\rangle} \propto \frac{d\rho(x,y)}{\langle \rho\rangle}$. Combining these relations (with $n(x,y) = n_0 + dn(x,y)$), we obtained:

$$\frac{dn(x,y)}{n_0} \propto \frac{dI(x,y)}{\langle I\rangle} \propto \frac{d\rho(x,y)}{\langle \rho\rangle}.$$

The resulting optical lattice was modeled with the Anderson-disordered tight-binding model (TBM) Hamiltonian. We assumed a single optical state (photon mode) per lattice site and nearest-neighbor hopping. The Hamiltonian is given by:

$$H = \sum_i \varepsilon_i \mid i\rangle\langle i \mid +t \sum_{\langle ij\rangle} (\mid i\rangle\langle j \mid + \mid j\rangle\langle i \mid),$$

where $\varepsilon_i$ is the onsite energy at the lattice site $i$ , $| i\rangle$ and $| j\rangle$ are optical wavefunctions at sites $i$ and $j$ , $\langle ij\rangle$ denotes nearest neighbors, and $t$ is the overlap (hopping) integral between adjacent sites. The onsite potential $\varepsilon_i$ is expressed in terms of refractive index or intensity fluctuations:

$$\varepsilon_i \propto \frac{dn}{n_0} \propto \frac{dI}{\langle I\rangle}.$$

Thus, the Hamiltonian directly incorporates structural disorder extracted from optical intensity variations. Structural disorder was quantified from the ensemble-averaged inverse participation ratio (IPR) of the eigenfunctions of the optical lattice:

$$\langle IPR\rangle_{L\times L} = \frac{1}{N}\sum_{i=1}^{N}\int_0^L\int_0^L E_i^4(x,y)dxdy$$

where $E_i$ is the $i$-th eigenfunction of the optical lattice, $N = (L/a)^2$ is the total number of eigenfunctions (with lattice spacing $a = dx = dy$), and the averaging is performed over all eigenfunctions.

The IPR measures the degree of localization and therefore structural disorder. For weakly disordered biological media, the ensemble average IPR scales as:

$$STD(IPR) \sim L_d \sim \sigma(dn) \times l_c,$$

where $\sigma(dn)$ is the magnitude (standard deviation) of refractive index fluctuations, and $l_c$ is the spatial correlation length [15,17–19].

***Results:*** *IPR-light-localization applications:*

We applied the IPR-light-localization technique to four types of cancer tissue samples (lung, breast, colon, and pancreatic) at various stages to assess its performance and universality of the optical parameter, <*IPR*>, in cancer detection.

*Sample collection*: Automated transmission microscopy was used to collect a large volume of images. In particular, images were acquired using a motorized Olympus BX61 microscope equipped with a CCD camera and PRIOR Test Control (OptiScan software), with auto-scan and autofocus features, as shown in the cartoon in Fig.1. Imaging was performed using a 40x (UIS2) objective in transmission mode. A custom GUI-controlled data acquisition system was used. Thin sections 5 μm thick and 1.5 mm wide were imaged with an automated optical microscope.

Each approximately 1.5 mm sample area was divided into segments according to magnification, yielding ~300 micrographs per sample for analysis [20,21]. Commercially available tissue microarray (TMA) samples from different cancer types, including lung, breast, colon, and pancreatic (with 24-32 tissue cores per slide; for control and Stages I-III), were obtained from BioMax, TriStar Technology, USA; Biotechne, USA; BioChain, USA, etc. These tissue samples were mounted on glass slides for the optical microscopy experiments [22–24]. These tissues were obtained for label-free and stained tissue samples.

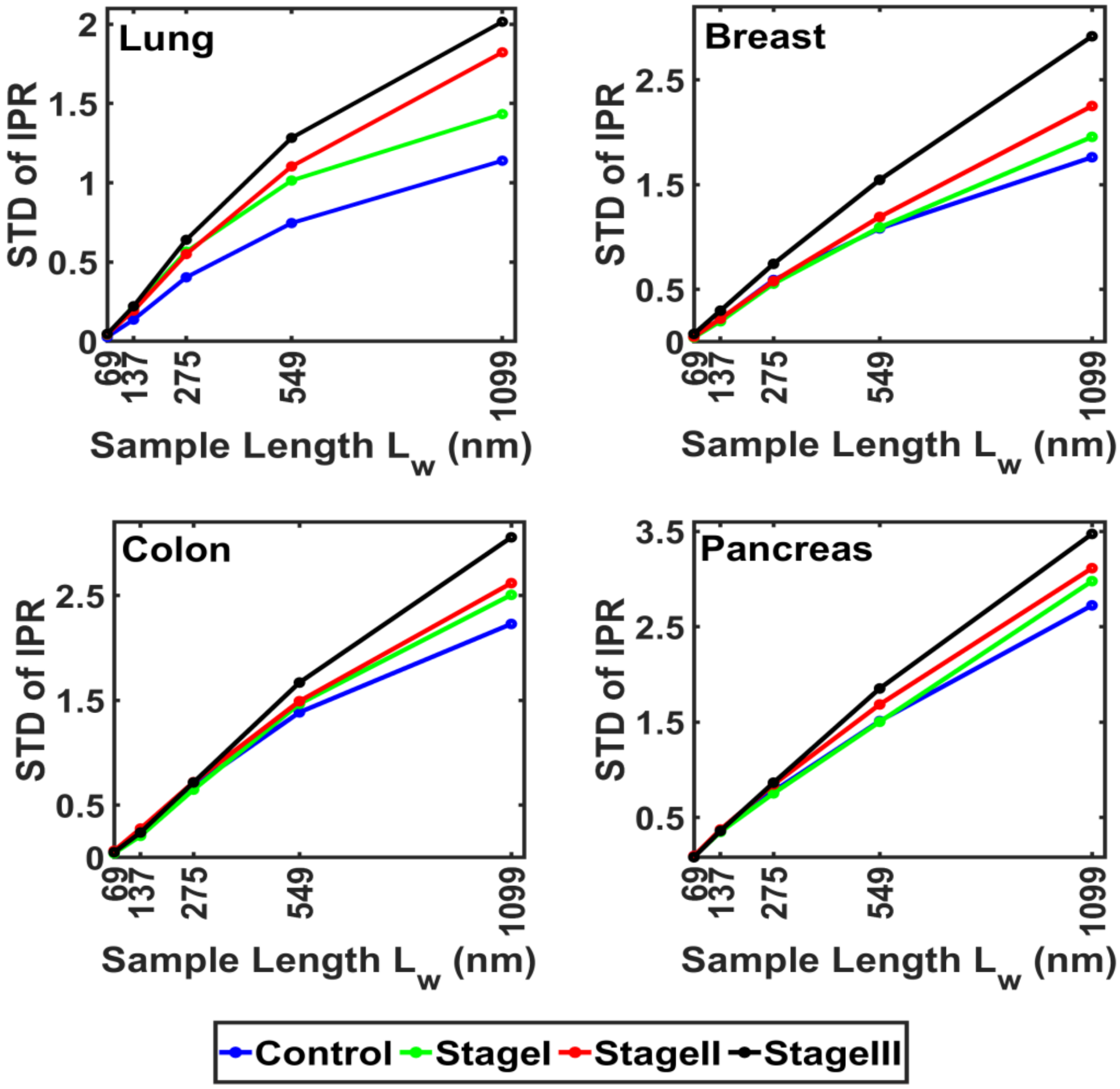

Fig. 2. The STD of IPR plotted as a function of optical lattice sample size. It shows that larger length scales result in higher structural disorder across all cancer types, including lung, breast, colon, and pancreatic cancers. (Student's t-test, $p < 0.05$. For each control and cancer stage, approximately 4–6 tissue cores from different patients and ~20 spots per core were considered; that is, on average, ~80-100 spots for each control and each stage (Stage1-III) were analyzed.)

---

*Tissue sample length-dependent IPR value:* We first examine the scale dependence of the IPR-derived structural disorder parameter using square optical lattice windows $L_w \times L_w$ of size pixels, where $L_w$ = 2, 4, 8, 16, and 32, and corresponding sample lengths are 69 nm, 137 nm, 275 nm, 549 nm, and 1099 nm. The average standard deviation of the ensemble-average IPR increased with the length for both the control and cancer stages. It has already been shown that in a two-dimensional uniform system, IPR($L$) increases from zero as L increases and saturates at 2.25. In a disordered system, it increases from zero and saturates at a value greater than 2.25. A larger length window contains a broader tissue region in each optical lattice and, as a result, exhibits higher mass density and refractive index fluctuations over a larger spatial extent. Cancer stages show higher

*STD(IPR(L))* for larger sample sizes ($L \times L$) compared to control, and this trend is observed across different cancer types, including lung, breast, colon, and pancreatic cancers studied here. Figure 2 also clearly indicates that the deviation from control to cancer occurred at a length scale of ~150 nm in optical microscopy images at 40x magnification, for all the cancer cases studied here.

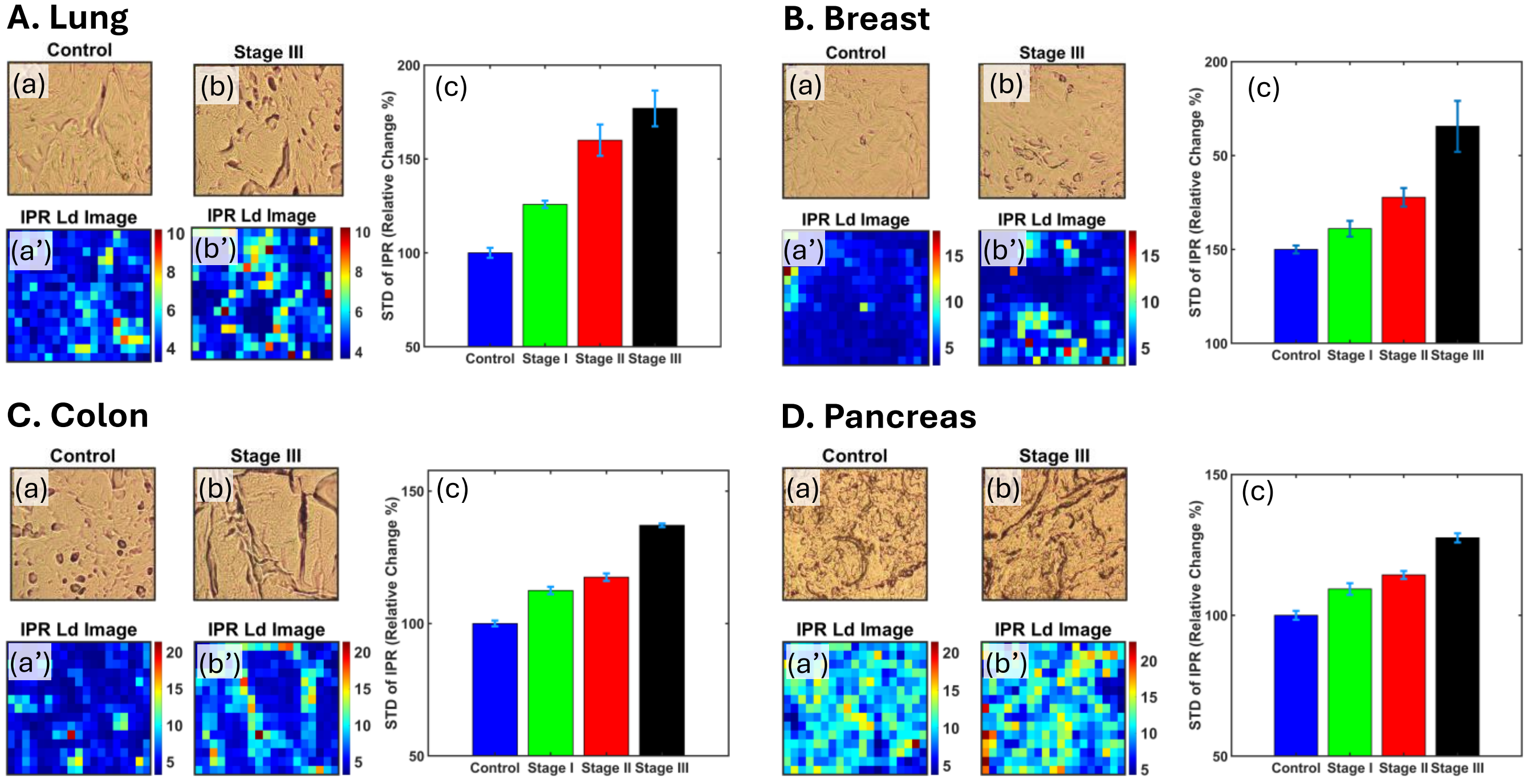

**Fig. 3.** Light-localization IPR analysis of (A) lung, (B) breast, (C) colon, and (D) pancreatic cancers. Bright-field images of (a) control and (b) Stage III tissues and their corresponding IPR maps (a′, b′) are shown. (c) Bar graphs show ⟨IPR⟩ for control and Stages I–III. ⟨IPR⟩ increased progressively with cancer stage, with a significant 27–77% increase in Stage III relative to control (at sample size: ~1.2 $\mu^2$). (Student's *t*-test, $p < 0.05$. For each control and cancer stage, approximately 4–6 tissue cores from different patients and ~20 spots per core were considered; that is, on average, ~80-100 spots for each control and stage (StageI-III) were analyzed.)

*IPR-light localization imaging and quantification:* To demonstrate the potential of this experimental and numerical optical technique, structural disorder in TMA samples from lung, breast, colon, and pancreatic cancers was quantified using transmission optical microscopy and IPR-light localization analysis. For each cancer type and stage (Control and Stages I–III), ~20-25 tissue microscopy images were analyzed. As presented in Fig. 3, representative brightfield images (a–b) and corresponding IPR color maps (a′–b′) are shown for lung (A), breast (B), colon (C), and pancreatic (D) tissues.

As shown in Fig. 3A–D(c), the STD of IPR progressively increased from Control to Stage III across all cancers. Relative to Control, the increases for Stages I, II, and III were 26%, 60%, and 77% in lung (IPR measured at a sample size L*L = 1099*1099 $nm^2$ ); 11.05%, 27.76%, and

65.54% in breast; 12.45%, 17.48%, and 37.08% in colon; and 9.4%, 14.3%, and 27.5% in pancreatic tissues, respectively. The increased *STD(IPR)* indicates greater spatial structural disorder, supporting IPR as a quantitative marker for cancer detection and reliable diagnostic differentiation. We have chosen *STD(IPR)*, which is a better parameter to deal with than *<IPR>*.

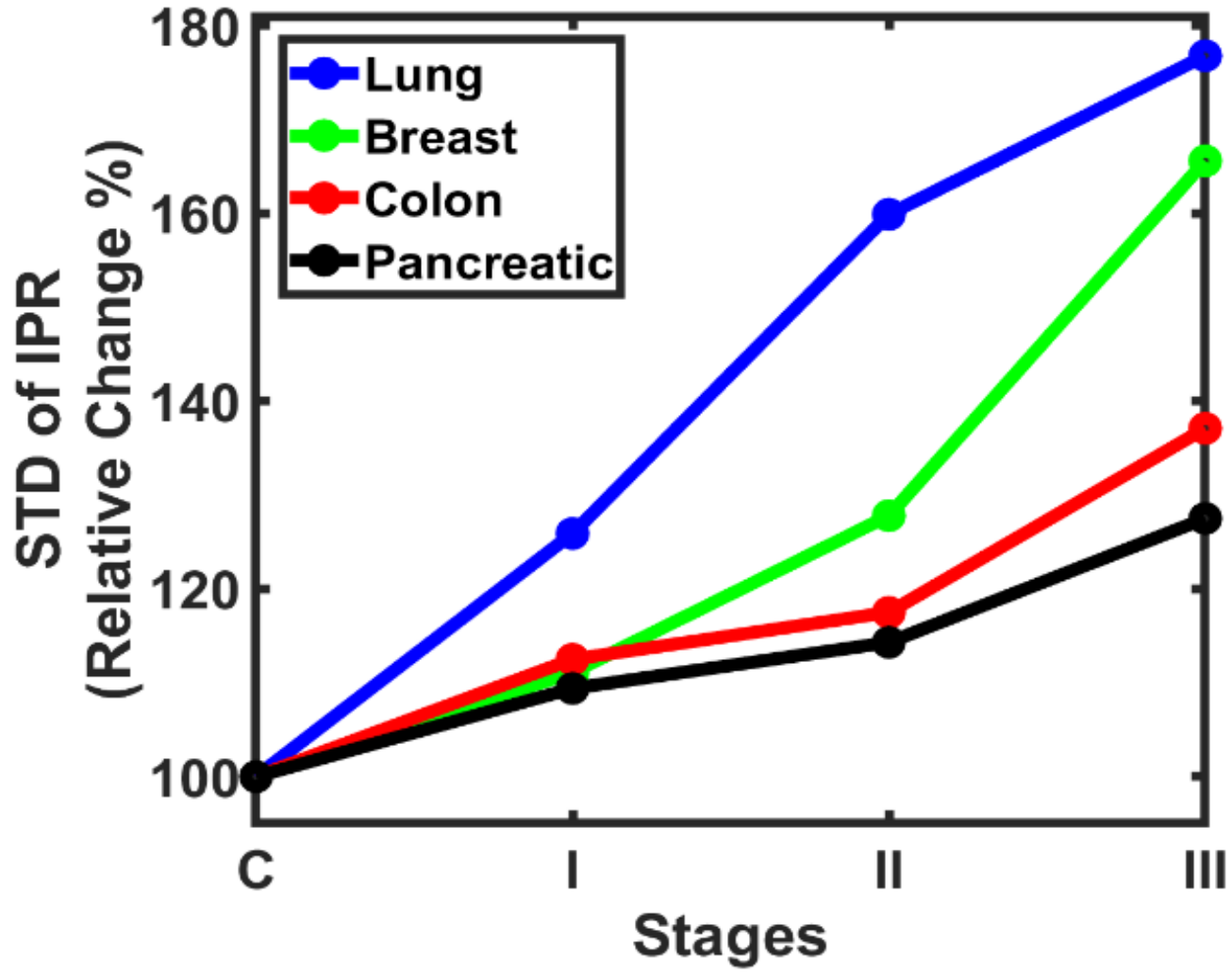

Fig 4. The standard deviation of the IPR value for the control (C) and stages I, II, and III shows an increasing trend for most cancer cases, rescaled to 100, for each control (C). (Sample size: ~1.2 $\mu^2$, error bars are as shown in Fig.3)

In Fig. 4, we show the progressive increase in the *STD (IPR)* across cancer stages for lung, breast, colon, and pancreatic cancers. The monotonic increase in the STD of IPR with cancer stage indicates that structural disorder increases with cancer progression across several cancer types. As a result, IPR analysis becomes more sensitive for quantifying progressive structural disorder and serves as a robust optical biomarker for cancer staging. The results also show the typical length scale at which one can detect IPR cancer progression by optical microscope.

*Cross-Platform Significance:* To our knowledge, this study is the first implementation of an IPR-based light-localization computational framework that uses conventional brightfield transmission optical micrographs of tissue samples for carcinogenesis or early cancer detection. A different imaging, technical, and algorithmic version of the method has previously been applied to transmission electron microscopy (TEM) to quantify nanoscale mass density fluctuations and to confocal fluorescence microscopy to measure molecular-specific nuclear structural disorder [15,18,25]. Although these imaging modalities provide high spatial resolution of tissue and cellular components at the nanoscale, they require DAPI-stained nuclear DNA samples and complex TEM sample preparation, including fixation, electron staining, ultrathin sectioning, and embedding, as well as specialized instrumentation for imaging. This makes them complex, time-consuming, and expensive imaging modalities.

The current IPR method for tissue imaging enables quantitative cancer staging and disorder assessment based on optical structural biomarkers, directly from lab-grade transmission optical microscope images, using a new, advanced, mesoscopic-based optical localization IPR quantification algorithm and steps. This optical microscopy instrument is common and accessible in pathology labs, unlike specialized systems such as TEM, confocal fluorescence microscopy, and spectroscopy. The IPR-based structural biomarker derived from these images can reduce equipment complexity, save time, and lower costs, thereby facilitating faster imaging and better compatibility with routine pathology slides. This approach supports analysis across key cancer types—lung, breast, colon, and pancreas—rather than being limited to a single cell line or disease model, and so universality.

Our optical platforms probe a common physical principle despite differences in resolution and contrast mechanisms; structural and macromolecular alterations in tissue associated with fluctuations in mass density and refractive index result in the spatial localization of optical eigenfunctions. The observed stage-dependent increase in single-parameter *STD(IPR)* is consistent with the increasing spatial heterogeneity of diseased tissues, driven by changes in chromatin structure, accumulation of intracellular macromolecular components, and uncontrolled cell growth. However, image-derived IPR may also be influenced by illumination, focus, camera response, sample staining, and image normalization. But our automated image scanning and autofocus reduce image acquisition problems, and ensemble-average IPR analysis on a large sample may improve statistical robustness. As a result, this approach could serve as a general optical computational framework for early cancer detection.

*Discussion and Conclusions:* We have established a mesoscopic physics–based light localization framework that quantitatively characterizes the degree of tissue structural disorder at the nano- to submicron scales by ensemble averaging across multiple tissues. By incorporating the Inverse Participation Ratio (IPR) into a unified diagnostic framework, this approach provides an objective, quantitative measure of structural heterogeneity and pixel-level disorder directly from standard transmission optical microscopy images. We demonstrate that tissue structural disorder quantified by ensemble-averaged <IPR>~$L_d = \sigma(dn) \times l_c$, increases approximately linearly with tumor/cancer stage across multiple cancer types. This biomarker provides more profound insights into cancer progression. Combined with statistical analysis, the IPR-based light localization approach enables sensitive, robust, and scalable cancer detection from routine microscopy images. The mesoscopic optical-physics-driven platform is designed for clinical integration, aiming to be cost-effective, improve diagnostic accuracy, and support early, personalized patient care.

## ACKNOWLEDGMENTS

The work was partially funded by NIH Grant number R21 CA260147